\documentclass[12pt]{article}

\usepackage{amsmath}
\usepackage{graphicx}
\usepackage{hyperref}
\usepackage[FIGTOPCAP]{subfigure}

\begin{document}

\title{Revisiting dissipative motion of a spinning heavy symmetric top and the rise of the top by friction}

\author{        V. Tanr{\i}verdi \\
tanriverdivedat@googlemail.com \\
Address: Bahad{\i}n Kasabas{\i} 66710 Sorgun-Yozgat TURKEY
}

\date{}

\maketitle

\begin{abstract}
	The dissipative motion and the rise of a heavy symmetrical top with a hemispherical peg are studied.
        A model taking the fixed point of the top as the center of the peg is considered when the top completely slips and the rolling motion is ignored.
        This is different from existing models like Jellet's one.
        Jellett's model and pure slipping are compared for different tops for the rise of the top.
\end{abstract}

\section{Introduction}

The motion of a heavy symmetric top with one point fixed is one of the interesting topics of classical mechanics.
In some cases, one can consider the fixed point as the tip of the top \cite{Tanriverdi_dissipative}.
And, for such considerations, the tip should be taken as a point.
On the other hand, in general, the top's tip is not completely fixed since the tip is not a point, and the top rolls and slips on the surface.
The rolling motion occurs due to rotations of the top, and the touchpoint can make some circle-like paths on the surface \cite{Fokker1941, Parkyn}.

One can consider two limiting cases for the motion of the top by accepting there is not any initial translational motion.
In one case, there is not any slipping, and the top completely rolls on the plane.
Rolling motion takes place when the spin angular velocity is not high and friction is not too low.
This rolling motion can take place together with a periodic slipping \cite{Parkyn}.
And, in general, there is not any fixed point when the rolling motion is present.
In the other limiting case, the top completely slips on the surface, and it always touches at the same point on the surface.
Contrary to rolling motion, slipping takes place if the friction is low enough and the spin angular velocity is high enough.
For the completely slipping case, the fixed point of the top can be taken as the center of the peg when the bottom of the peg is hemispherical.
We should note that the radius of the peg also affects rolling and slipping conditions since it changes the velocity of the touchpoint.
Studying the motion of the top with a hemispherical peg is a previously used method \cite{Parkyn, Routh, Jellett, Yogi}.

Dissipative effects are important to explain the motion of the top in daily life.
In previous work, we considered air dissipation and friction at the touchpoint \cite{Tanriverdi_dissipative}, and that model does not give the rise of the top.
At that work, one can find a short summary of various works related to dissipative effects.

In most of the situations that we encounter in daily life, friction results in the ceasing of motion.
On the contrary, for the motion of a heavy symmetric top, friction does something unintuitive: It rises the top.
According to Gray, the first statement that the rise occurs as a result of slipping is given by Smith \cite{Gray}.
Jellett explicitly gave equations for the rise of the top without explicitly defining reaction force \cite{Jellett}.
And, in that work, the center of mass is considered as fixed, and by using limit, it is shown that slipping causes an increase in the precession angular velocity which results in the rise.
Perry considered different observations related to the top and gave verbal explanations to the rise by considering Jellett's model \cite{Perry}.

Later various scientists have studied the rise of the top.
Fokker observed that the rise time is shorter for greater radii which is consistent with Jellett's model \cite{Fokker1941}.
Hugenholtz gave an explanation to the rise of the top by "rolling friction" without writing equations explicitly \cite{Hugenholtz}.
Braams considered "sliding friction" and its effect on the rise, and he stated that "sliding friction" contributes to the rise of the top in the fast precession, and he also stated that the "rolling friction" contributes to the rise without giving explicit equations \cite{Braams}.
Parkyn has considered the change of reaction force and change of the center of mass with respect to ground and given related equations for the motion of the top including the rise term \cite{Parkyn}.
Yogi has defined new angular velocities to avoid singularities and considered the change in the reaction force, and he has solved numerically resulting equations giving the rise of the top \cite{Yogi}.
Moffatt et al. have considered a new reference frame which is a mixed one of stationary and body reference frames to study the motion of symmetric rigid bodies without considering the change of reaction force \cite{Moffatt}. 
We should note that works of Jellett, Parkyn, Yogi and Moffatt et al. consider the top's center of mass as fixed, and the rising term is a function of the distance between the radius of the peg and the center of mass \cite{Parkyn, Jellett, Yogi, Moffatt}.

Hugenholtz and Braams stated that "rolling friction" contributes to the rise of a heavy symmetric top, however, their statement is vague since they did not give any explicit equations.
All existing explicit models related to the rise of a heavy symmetric top are based on the friction at the touchpoint occurring as a result of slipping.
Some authors have used "sliding friction" to mention this, and Parkyn used slipping to describe it.
We will use slipping friction like Parkyn since some people can understand "sliding friction" as ordinary sliding, which is not the cause of the rise.

In general, there is not any fixed point of a heavy symmetric top.
There is a drawing in Perry's work showing the center of mass of the top is fixed while rolling motion takes place, i.e. figure 32 \cite{Perry}.
On the other hand, one can see from a video for a rising top, available as supplementary material in Cross' experimental work \cite{Cross2013}, that as a result of rolling motion, the center of mass of the top is not fixed like Perry's drawing.
From that video, one can also see that the top also nutates and precesses.
And, the position of the center of mass changes due to these besides rolling motion.
However, we should note that there are some cases in which motion takes place similar to Perry's drawing, and in some of these cases, the position of the center of mass may not change.
But, this does not describe the motion properly for all cases.
Then, it is possible to say that considering the center of mass as the fixed point is not true in general.
On the other hand, the radial center of the peg does not change during the nutation and precession, and one can accept it as the fixed point if the rolling motion is ignored.
And, in this work, we will consider the fixed point of the top as the radial center of the peg and rotations around it, assuming rolling motion is not present.

Another difference with previous works is that they take the reaction force as torque affecting the motion of the top which is not in this work.
This is related to the choice of the fixed point.
We already mentioned that in some of the previous works, the reaction force is different from mass times gravitational acceleration \cite{Parkyn, Yogi}.
On the other hand, Quinn and Picard measured the mass change of a gyroscope during the rotation and find "no dependence on speed or sense of rotation" in their experiment \cite{QuinnPicard}. 
We should note that Quinn and Picard used a gyroscope with casings which is different from the symmetric top.
Nevertheless, in this work, we will assume that the reaction of the surface is equal to mass times gravitational acceleration.

In section \ref{two}, we will derive equations defining the motion of the top with a hemispherical peg when the rolling motion is not present and the fixed point is the center of the hemispherical peg.
We will include air dissipation and slipping friction at the touchpoint.
In section \ref{three}, we will review Jellett's model.
In section \ref{four}, we will numerically solve obtained equations for a dissipative experimental situation and two hypothetical tops to study the rise, and then we will conclude in section \ref{five}.

\section{Pure slipping model}
\label{two}

\begin{figure}[!h]
        \begin{center}
		\subfigure[]{\includegraphics[width=5.5cm]{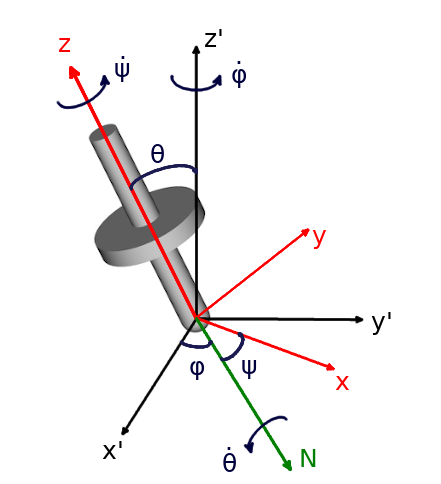}}
		\subfigure[]{\includegraphics[width=5.5cm]{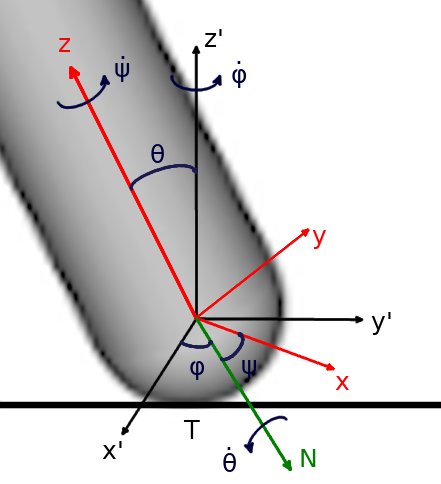}}
		\caption{Heavy symmetric top, stationary reference frame ($x',y',z'$), center of peg-body reference frame ($x,y,z$), line of nodes $N$, Euler angles ($\theta, \phi, \psi$), angular velocities ($\dot \theta, \dot \phi, \dot \psi$) and the touchpoint T. a) General view. b) Peg of the top.
                }
        \label{fig:hst}
        \end{center}
\end{figure}

In figure \ref{fig:hst}(a), one can see a symmetric top, body reference frame $(x,y,z)$ and stationary reference frame $(x',y',z')$ together with line of nodes $N$.
Origins of the reference frames are placed to the center of the hemispherical peg since it is taken as the fixed point by considering pure slipping and ignoring rolling motion.
In some other models, a body reference frame whose origin is at the center of mass can be used.
We will use the name the center of peg-body reference frame to emphasize the difference from that type of model.
The hemispherical peg touches the surface at point $T$ shown in figure \ref{fig:hst}(b).
As the top rotates, the position of the peg's center and the point $T$ on the surface do not change.
However, the point on the peg touching the surface changes.

Rotations of the symmetric top ($I_y=I_x$) can be described by Euler equations which can be written as
\begin{eqnarray}
       \tau_x&=&I_x \dot w_x+w_y w_z (I_z-I_x),  \nonumber \\
       \tau_y&=&I_x \dot w_y+w_x w_z (I_x-I_z), \label{eeq} \\
       \tau_z&=&I_z \dot w_z, \nonumber
\end{eqnarray}
where $\tau_i$, $I_i$ and $w_i$ correspond to $i^{th}$ component of torque, moments of inertia and angular velocity in the body coordinate system, respectively.
In terms of Euler angles, angular velocities can be written as
\begin{eqnarray}
        w_x&=&\dot \theta \cos \psi+\dot \phi \sin \theta \sin \psi, \nonumber \\
        w_y&=&-\dot \theta \sin \psi+ \dot \phi \sin \theta \cos \psi, \label{ang_vel}\\
        w_z&=&\dot \psi+\dot \phi \cos \theta, \nonumber 
\end{eqnarray}
where $\dot \theta$, $\dot \phi$ and $\dot \psi$ are nutation, precession and spin angular velocities, respectively.
These can be seen in figure \ref{fig:hst}.

By considering pure slipping, one can obtain the slipping velocity of the touchpoint of the top by using $\vec v=\vec w \times \vec r=\vec w \times (-R_p \hat z')$ as
\begin{eqnarray}
	\vec v=&R_p &[(\dot \theta \cos \theta \sin \psi+\dot \psi \sin \theta \cos \psi)\hat x+(\dot \theta \cos \theta \cos \psi-\dot \psi \sin \theta \sin \psi)\hat y \nonumber \\
	& &+(-\dot \theta \sin \theta)\hat z],
	\label{slipping_velocity}
\end{eqnarray}
where $R_p$ is the radius of the hemispherical peg.
From this equation, it can be seen that, as expected, the slipping velocity is independent of the precession angular velocity $\dot \phi$ which takes place on the vertical axis passing through the center of the peg and the touchpoint.
The friction should be in the reverse direction of the touch point's velocity and can be written as $\vec f=- k N \vec v / |\vec v|$, where $k$ is the positive friction constant, and the reaction force is taken as $N=Mg$.
Then, the torque due to this friction can be obtained by using $\vec \tau=\vec R_p \times \vec f$ as
\begin{eqnarray}
	\vec \tau= &\frac{k N R_p^2} {|\vec v|}&  [(-\dot \theta \cos \psi+\dot \psi \sin \theta \cos \theta \sin \psi)\hat x\nonumber \\
	& & \,+(\dot \theta \sin \psi+\dot \psi \sin \theta \cot \theta \cos \psi)\hat y+(-\dot \psi \sin^2 \theta)\hat z],
\end{eqnarray}
where $|\vec v|=R_p \sqrt{\dot \theta^2 +\dot \psi^2 \sin^2 \theta}$.

Air dissipation is another factor in the motion of the top, and characteristics of this dissipation for the spin angular velocity and nutation \& precession angular velocities are different.
Skin friction drag is mainly responsible for air dissipation of streamlined bodies, and pressure drag is mainly responsible for air dissipation of blunt bodies \cite{Anderson}.
These can depend on various factors, e.g. the density of the fluid, the cross-section of the body, relative velocity.
However, in this work, we will not go into details and use simple models. 
Skin friction drag is mainly responsible from air dissipation for the spin angular velocity, and this dissipative torque can be modelled as $\vec \tau_{\dot \psi}=(d_1 \dot \psi+d_2 \dot \psi^2) \hat z$.
Pressure drag is mainly responsible for nutation \& precession angular velocities, and corresponding dissipative torque can be written as : $\vec \tau_{\dot \theta}= c_1 \dot \theta (\cos \psi \hat x-\sin \psi \hat y)$ and $\vec \tau_{\dot \phi}= c_2 \dot \phi \hat z'= c_2 \dot \phi(\sin \theta \sin \psi \hat x+\sin \theta \cos \psi \hat y+\cos \theta \hat z)$.
Since the origin of dissipations due to nutation \& precession angular velocities are the same and the cross-sections are the same for both rotations, air dissipation coefficients for nutation \& precession angular velocities can be considered as the same, i.e. $c=c_1=c_2$.
Signs of $c$ and $d_1$ should be negative, and the sign of $d_2$ should be negative of the sign of $\dot \psi$. 

In the body coordinate system, the gravitational torque can be written as
\begin{equation}
        \vec \tau_g=-Mg \tilde l \sin \theta (-\cos \psi \hat x+\sin \psi \hat y),
\end{equation}
where $\tilde l$ is the distance from the peg's center to the center of mass.

If one includes all of the mentioned torques in Euler equations, one can get the following equations
\begin{eqnarray}
        \ddot \theta&=& -\frac{I_z \dot \phi \sin \theta}{I_x}(\dot \psi+\dot \phi \cos \theta)+\dot \phi^2 \sin \theta \cos \theta+ \frac{Mg \tilde l }{I_x} \sin \theta+\frac{c \dot \theta}{I_x}-\frac{k M g R_p^2 \dot \theta}{I_x |\vec v|},  \nonumber \\
       \ddot \phi&=& \frac{I_z \dot \theta}{I_x \sin \theta}( \dot \psi +\dot \phi \cos \theta)- \frac{ 2 \dot \theta \dot \phi \cos \theta}{ \sin \theta} +\frac{c \dot \phi}{I_x}+\frac{k M g R_p^2 \dot \psi \cos \theta}{I_x |\vec v|},  \label{diffeqns} \\
	\ddot \psi&=& - \frac{I_z \dot \theta \cos \theta}{I_x \sin \theta} ( \dot \psi +  \dot \phi \cos \theta )+\frac{2 \dot \theta \dot \phi \cos^2 \theta}{\sin \theta} +\dot \theta \dot \phi \sin \theta +c \dot \phi \cos \theta\left(\frac{1} {I_z}-\frac{1} {I_x}\right)\nonumber \\
	& & +\frac{d_1 \dot \psi+d_2 \dot \psi^2}{I_z}-\frac{k M g R_p^2 \dot \psi }{|\vec v|}\left(\frac{\sin^2 \theta} {I_z}+\frac{\cos^2 \theta} {I_x}\right), \nonumber
\end{eqnarray}
where moments of inertia should be calculated by considering that rotations take place around the peg's center.
These equations describe rotations of the heavy symmetric top when the center of the peg is fixed, air dissipation is included according to mentioned models, the touchpoint of the peg slips on the ground and the rolling motion is not present.
The term $k M g R_p^2 \dot \psi \cos \theta/(I_x |\vec v|)$ is the rise term of this model.
When dissipative terms are set to zero, these equations become the same as the ones obtained from Lagrangian \cite{Tanriverdi2019a}.

One can simplify this model by ignoring air friction and friction at the touchpoint due to motion in $\theta$ since $|\dot \theta| << |\dot \psi|$ and obtain
\begin{eqnarray}
       \ddot \theta&=& -\frac{I_z \dot \phi \sin \theta}{I_x}(\dot \psi+\dot \phi \cos \theta)+\dot \phi^2 \sin \theta \cos \theta+ \frac{Mg \tilde l }{I_x} \sin \theta,  \nonumber \\
       \ddot \phi&=& \frac{I_z \dot \theta}{I_x \sin \theta}( \dot \psi +\dot \phi \cos \theta)- \frac{ 2 \dot \theta \dot \phi \cos \theta}{ \sin \theta} +\frac{k M g R_p^2 \dot \psi \cos \theta}{I_x |\vec v|},  \label{diffeqns_odp} \\
	\ddot \psi&=& - \frac{I_z \dot \theta \cos \theta}{I_x \sin \theta} ( \dot \psi +  \dot \phi \cos \theta )+\frac{2 \dot \theta \dot \phi \cos^2 \theta}{\sin \theta}+\dot \theta \dot \phi \sin \theta \nonumber \\ 
       & & -\frac{k M g R_p^2 \dot \psi }{ |\vec v|}\left(\frac{\sin^2 \theta} {I_z}+\frac{\cos^2 \theta} {I_x}\right). \nonumber
\end{eqnarray}
One can see that all dissipative effects are related to motion in $\dot \psi$ in these equations, and they originate from the friction at the touchpoint.

\section{Jellett's model for the rise of the top}
\label{three}

In this part, we will write Jellett's model in terms of the parameters of this work.
We should note that Jellett used an upside-down center of mass-body reference frame, and for simplicity ignored air dissipation and frictions with respect to $\dot \theta$ and $\dot \phi$.
We will write Jellett's parameters inside square parentheses to eliminate confusion.

Jellett gave equations for the rise of the top in equations (4) and (5) in his work \cite{Jellett}.
Jellett's parameters and variables can be written in terms of parameters and variables of this work as: $[\theta]=\theta$, $[\psi]=\phi$, $[\phi]=-\psi$, $[R \cos \epsilon]=N$, $[\xi]=R_p \sin \theta$, $[\eta]=\tilde l+R_p \cos \theta$, $[z']=\tilde l \sin \theta$, $[z]=R_p+\tilde l \cos \theta$ and $[R \sin \epsilon]=-f$.
Due to the usage of the upside-down center of mass-body reference frame, the spin angle and friction force get a minus sign.
In Jellett's work, the reaction force is not specified, and we will take $N=M g$.
Then, one can obtain equations of motion for Jellett's model in terms of parameters and variables of this work by considering $f=N k$ as  
\begin{eqnarray}
       \ddot \theta&=& -\frac{\tilde I_z \dot \phi \sin \theta}{\tilde I_x}(\dot \psi+\dot \phi \cos \theta)+\dot \phi^2 \sin \theta \cos \theta+ \frac{M g \tilde l }{I_x} \sin \theta,  \nonumber \\
       \ddot \phi&=& \frac{\tilde I_z \dot \theta}{\tilde I_x \sin \theta}( \dot \psi +\dot \phi \cos \theta)- \frac{ 2 \dot \theta \dot \phi \cos \theta}{ \sin \theta} + \frac{M g k (\tilde l + R_p \cos \theta)}{\tilde I_x \sin \theta},  \label{jdiffeqns} \\
	\ddot \psi&=& - \frac{\tilde I_z \dot \theta \cos \theta}{\tilde I_x \sin \theta} ( \dot \psi +  \dot \phi \cos \theta )+\frac{2 \dot \theta \dot \phi \cos^2 \theta}{\sin \theta}+\dot \theta \dot \phi \sin \theta  \nonumber \\ & & 
        -\frac{M g k \cos \theta (\tilde l + R_p \cos \theta)}{\tilde I_x \sin \theta} - \frac{M g k R_p \sin \theta}{\tilde I_z}. \nonumber
\end{eqnarray}
We have used $\tilde I_i$ to emphasize that moments of inertia should be calculated in the center of mass-body reference frame.
In Jellett's model the rise term is $M g k (\tilde l + R_p \cos \theta)/(\tilde I_x \sin \theta)$ which linearly depends on $\tilde l$.
It can be seen that this is different from the rise term obtained by considering pure slipping.
We should note that signs of dissipative terms should be determined by considering the sign of $\dot \psi$.

\section{Numerical solutions}
\label{four}

In this section, we will numerically solve different cases with studied models.
Firstly, we will numerically solve a previously experimentally studied case with the pure slipping model, which is figure 5 in Cross' work \cite{Cross2013}.
This case is also studied with a different model, previously \cite{Tanriverdi_dissipative}.
To get consistent results with the figures given in the experimental work, we will take $\dot \phi$ and $\dot \psi$ as negative similar to previous work.
Then, we will study the rise of the top with the dissipation related parameters which are obtained by considering the experimental case.
We will also consider the simplified pure slipping model to get a better comparison with Jellett's model since air dissipation and friction at the touchpoint due to motion in $\theta$ are ignored in these two models.

\subsection{Dissipative motion}

By using given values in the experimental work for figure 5, parameters of the top becomes $M=105\, gr$, $\tilde l=20.9\, mm$, $R_p=0.1 mm$, $I_x=8.44 \times 10^{-5}\, kg \, m^2$ and $I_z=7.23 \times 10^{-5}\, kg \, m^2$ where moments of inertia are calculated according to the peg's center.
The initial values are taken as $\theta_0=0.117 \, rad$, $\dot \theta_0=2.23 \, rad \, s^{-1}$, $\dot \phi_0=-17.1 \, rad \, s^{-1}$ and $\dot \psi_0=-126 \, rad \, s^{-1}$ which are the same with previous model.
To get consistent results with experimental results, dissipative constants are taken as: $k=0.03 $, $c=-6.3 \times 10^{-6} \, kg \, m^2\, s^{-1} $, $d_1=-1.1 \times 10^{-6} \, kg \, m^2\, s^{-1}$, $d_2=-sgnm (\dot \psi) 1.1\times 10^{-8} \, kg \, m^2$.
These constants are not the same as the previous model since models are different.
The gravitational acceleration is taken as $g=9.81 \, m\, s^{-2}$.

\begin{figure}[h!]
        \begin{center}
	\subfigure[]{
                \includegraphics[width=2.8cm,height=2.8cm]{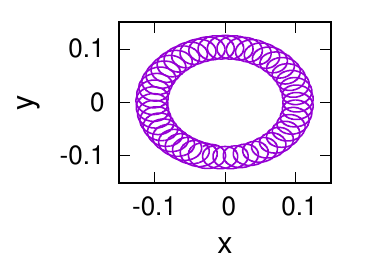}
                 }
        \subfigure[]{
                \includegraphics[width=2.8cm,height=2.8cm]{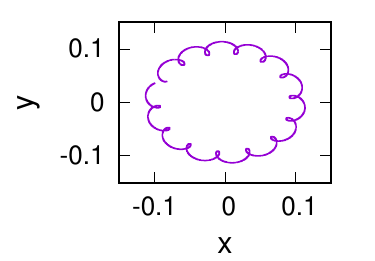}
                 }
        \subfigure[]{
                \includegraphics[width=2.8cm,height=2.8cm]{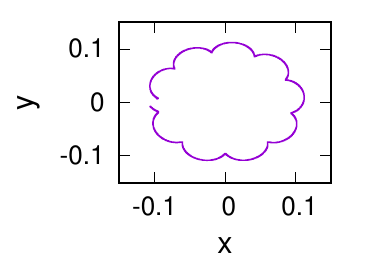}
                 }
        \subfigure[]{
                \includegraphics[width=2.8cm,height=2.8cm]{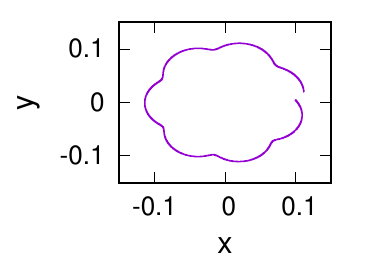}
                 }
\caption{Projections of shapes for the locus of the figure axis.
                 Time intervals (in seconds) for figures as follows: (a) $[0.73,3.45]$, (b) $[18.00,19.48]$ (c) $[23.60,24.87]$ (d) $[27.80,28.86]$.
                 Initial values are $\theta_0=0.117\,rad$, $\dot \theta_0=-2.23\,rad\,s^{-1}$, $\dot \phi_0=-17.1 \,rad\,s^{-1}$ and $\dot \psi_0=-126 \,rad\,s^{-1}$.
                 }
        \label{fig:prj_fig5}
        \end{center}
\end{figure}

Results of numerical solutions for $\theta$ and angular velocities can be seen in figure \ref{fig:thetaphipsi_c} in the appendix.
Projections of shapes for the locus can be seen in figure \ref{fig:prj_fig5}.
One can see that these are very similar to the experimental results.

We should note that there are some differences between the results of the previous model and this model.
Fluctuations in $\theta$ and angular velocities are larger in the previous model, however, they are damped by the dissipative factors at the end of 31 seconds.
On the other hand, they are not damped in this model.
The main reason for this should be related to the constant friction term in the previous model, which is considered for the friction at the touchpoint.
In this model, this friction shows itself as functions of $\dot \theta/|\vec v|$ and $\dot \psi \sin \theta /|\vec v|$ in torque, and $\dot \theta/|\vec v|$ is small and $\dot \psi \sin \theta /|\vec v|$ is close to one.
Then, the fluctuations in $\theta$ are damped more slowly in this model.
We should also mention that there is a slight rise in the average value of $\theta$ which can be seen in figure \ref{fig:thetaphipsi_c}(a) in the appendix. 
Such a rise is not observed in the previous model.
However, it is hard to determine this slight rise without taking average because of the high fluctuations.
On the contrary, an increase of $\theta$ is observed in the experiment, which is not observed in this and previous models.
Nevertheless, these show themselves with some small differences in the shapes for the locus.
And, the results of both models for the shapes for locus are similar to the experiment.

\subsection{Rise of the top}
\label{rott}

In this part, we will first consider the rise of the top with a hypothetical top that is slightly different from the top considered in the previous example.
We will add this hypothetical top a hemispherical peg with radius $R_p=6.0\, mm$.
The mass of the top will be taken $M=110 \,gr$, and the distance between the center of mass and center of the peg will be taken as $\tilde l_1= 20 \,mm$.
We need moments of inertia in two reference frames which will be taken as follows $I_{x}=8.52 \times 10^{-5} \,kg\,m^2$ and $I_{z}=7.25\times 10^{-5} \,kg\,m^2$ in the center of peg-body reference frame, and $\tilde I_{x}=4.08 \times 10^{-5} \,kg\,m^2$ and $\tilde I_{z}=7.25\times 10^{-5} \,kg\,m^2$ in the center of mass-body reference frame.

We will numerically solve a case by the pure slipping and Jellett's models for this top.
Then, we will consider another hypothetical top and compare the pure slipping, the simplified pure slipping and Jellett's models.
We will cut numerical solutions when the top reaches nearly upright position, i.e. $\theta=0.02\,rad$, due to infinities at $\theta=0$.
We will consider configurations giving regular precession initially to reduce the fluctuations with two exceptions which will be explained later.

\subsubsection{Pure slipping}

In this part, we will study the rise of the mentioned hypothetical top by using equations \eqref{diffeqns}. 
We will take initial values as follows $\theta=0.5\,rad$, $\dot \theta=0$, $\dot \phi_0=-1.49\,rad\,s^{-1}$ and $\dot \psi_0=-200\,rad\,s^{-1}$.

\begin{figure}[h!]
        \begin{center}
        \subfigure[$\theta$]{
                \includegraphics[width=4.2cm]{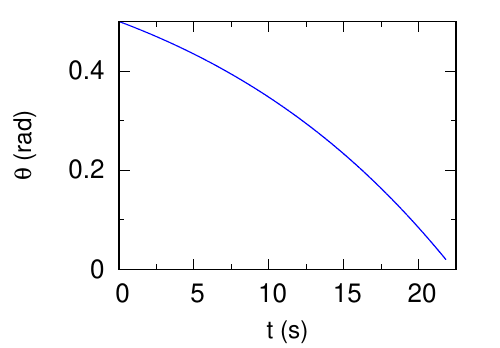}
                }
        \subfigure[$\dot \theta$]{
                \includegraphics[width=4.2cm]{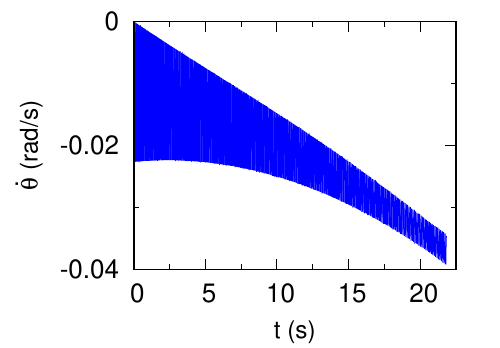}
                }

	\subfigure[$\dot \phi$]{
                \includegraphics[width=4.2cm]{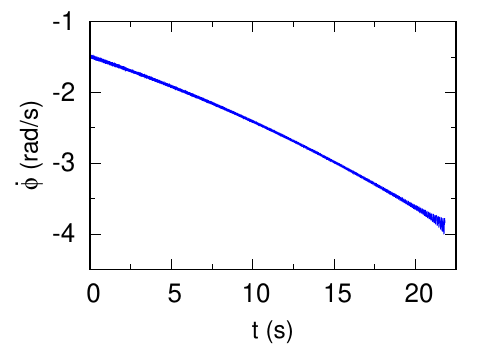}
                }
        \subfigure[$\dot \psi$]{
                \includegraphics[width=4.2cm]{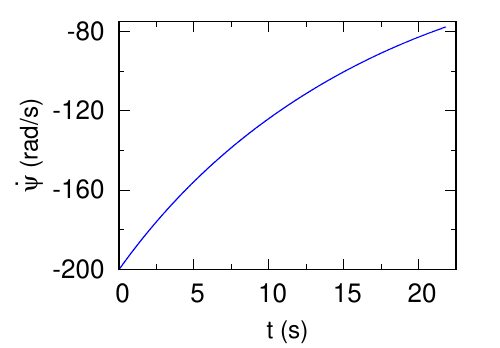}
                }
		\caption{Results of the numerical solution for $\theta$ (a), $\dot \theta$ (b), $\dot \phi$ (c) and $\dot \psi$ (d). Initial values: $\theta_0=0.5\,rad$, $\dot \theta_0=0$, $\dot \phi_0=-1.49 \,rad\,s^{-1}$ and $\dot \psi_0=-200 \,rad\,s^{-1}$. The parameters of the top are given in section \ref{rott}.
                }
        \label{fig:thetaphipsi_d}
        \end{center}
\end{figure}

Results of the numerical solution for $\theta$, $\dot \theta$, $\dot \phi$ and $\dot \psi$ can be seen in figure \ref{fig:thetaphipsi_d}.
One can see from figure \ref{fig:thetaphipsi_d}(a) that the top rises, and it comes nearly vertical position in $21.8$ seconds.
It can be seen from figure \ref{fig:thetaphipsi_d}(b) that there is a negligible nutation, which does not affect the motion much.
It can be seen from figure \ref{fig:thetaphipsi_d}(d) that the magnitude of $\dot \psi$ decreases which is an expected result of dissipation.
Its decrease is slightly faster in the beginning, and the main reason for this is the air dissipation.
It changes from $-200\, rad\,s^{-1}$ to $-77\, rad\,s^{-1}$ during the rise.
It can be seen from figure \ref{fig:thetaphipsi_d}(c) that the magnitude of $\dot \phi$ increases which is the result of the rise term originating from the dissipative torque at the touchpoint.
Its average value changes from $-1.49 \, rad \, s^{-1}$ to $-3.9 \, rad \, s^{-1}$, and there are some fluctuations as a result of nutation.

\begin{figure}[h!]
        \begin{center}
	        \subfigure[]{\includegraphics[width=4.2cm]{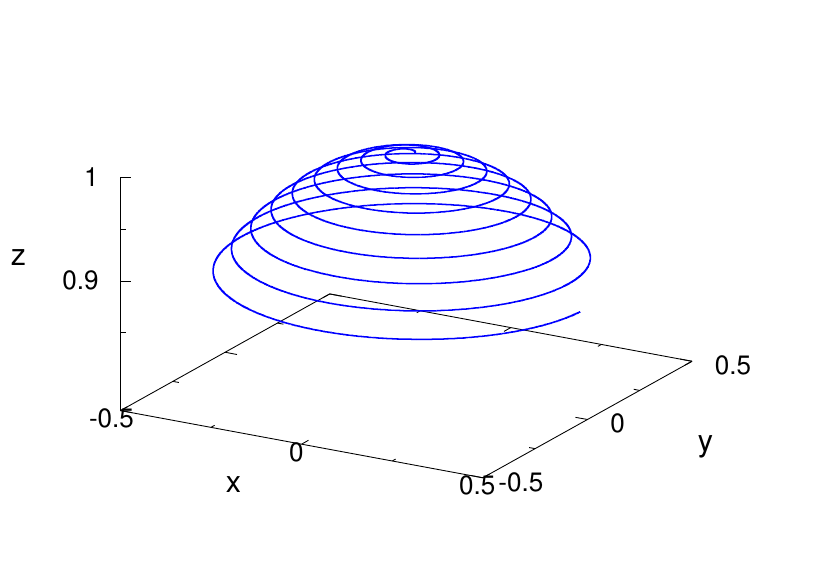}}
	        \subfigure[]{\includegraphics[width=4.2cm]{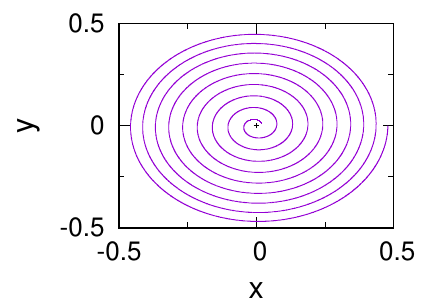}}
		\caption{Shapes for the locus (a) and its projection on to $xy$-plane (b) for the rise of the top.
		Initial values are given in figure \ref{fig:thetaphipsi_d}, and the solution is obtained by considering the pure slipping model.
		} 
        \label{fig:tt_prj_fig8}
        \end{center}
\end{figure}

In figure \ref{fig:tt_prj_fig8}, one can see the shapes for the locus during the rise and its projection onto $xy$-plane.
A spiral structure is observed during the rise.
We should note that this motion is different from "spiraling motion" which can be seen when conserved angular momenta are equal to each other \cite{Routh, Tanriverdi_abeql}.

\subsubsection{Jellett's model}

In this part, we will consider the rise of the top with Jellett's model, i.e. equations \eqref{jdiffeqns}.
Moments of inertia obtained by considering the center of mass-body reference frame will be used, and initial values will be the same as the previous solution except $\dot \phi_0$ which should be changed slightly in order to get a regular precession at the beginning, i.e. $\dot \phi_0=1.48 \,rad\,s^{-1}$.

\begin{figure}[h!]
        \begin{center}
        \subfigure[$\theta$]{
                \includegraphics[width=4.2cm]{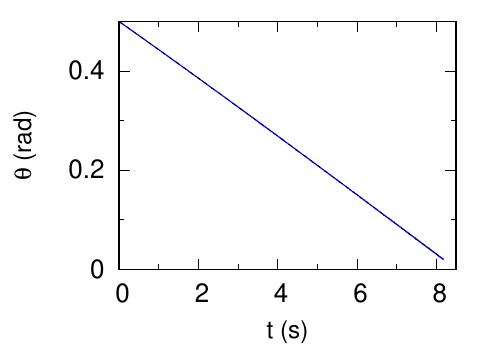}
                }
        \subfigure[$\dot \theta$]{
                \includegraphics[width=4.2cm]{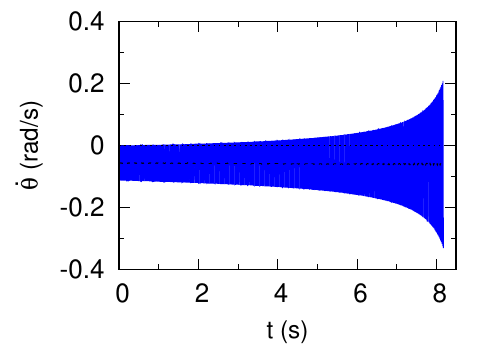}
                }

        \subfigure[$\dot \psi$]{
                \includegraphics[width=4.2cm]{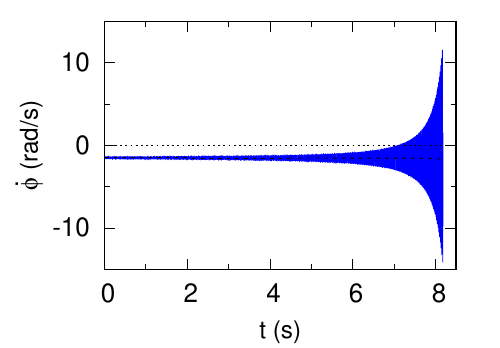}
                }
        \subfigure[$\dot \phi$]{
                \includegraphics[width=4.2cm]{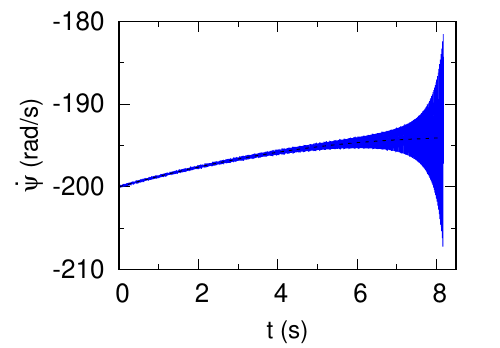}
                }
                \caption{Results of the numerical solution by considering Jellett's model for $\theta$ (a), $\dot \theta$ (b), $\dot \psi$ (c) and $\dot \phi$ (d). 
		Continuous (blue) lines show results of numerical solutions, and dashed (black) lines show averages.
		Initial values: $\theta_0=0.5\,rad$, $\dot \theta_0=0$, $\dot \phi_0=-1.48 \,rad\,s^{-1}$ and $\dot \psi_0=-200 \,rad\,s^{-1}$.
                }
        \label{fig:thetaphipsi_e}
        \end{center}
\end{figure}

The results of the numerical solution can be seen in figure \ref{fig:thetaphipsi_e}.
There are some changes from the previous results.
It can be seen that the top rises in $8.18$ seconds which is much faster than the previous model, and it is the result of $\tilde l$ dependent rise term.
Fluctuations of $\dot \theta$ increase as time passes, which are nearly ten times more than the pure slipping model.
But, the average value of $\dot \theta$ does not change much, and it is around $-1.6\,rad\,s^{-1}$.
Fluctuations of $\dot \phi$ also increase, and its average changes from $-1.48\,rad\,s^{-1}$ to $-1.6\,rad\,s^{-1}$ which is less than previous model.
$\dot \psi$ also fluctuates toward the end of the rise and its average becomes  $-194\,rad\,s^{-1}$ whose magnitudes are much more than the one in the pure slipping as a result of ignoring air dissipation.
The growth of fluctuations occurs due to ignoring air dissipation and fluctuations in relative small values of $\theta$.

\begin{figure}[h!]
        \begin{center}
                \subfigure[]{\includegraphics[width=4.2cm]{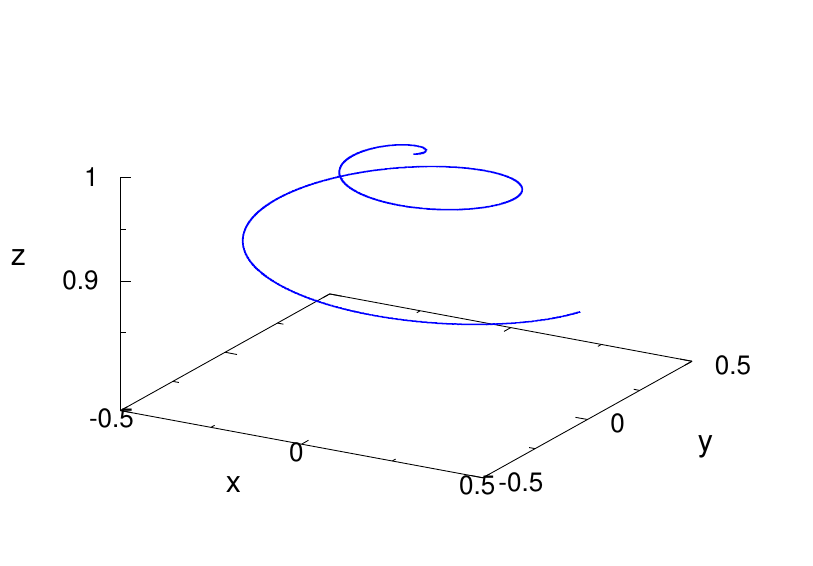}}
                \subfigure[]{\includegraphics[width=4.2cm]{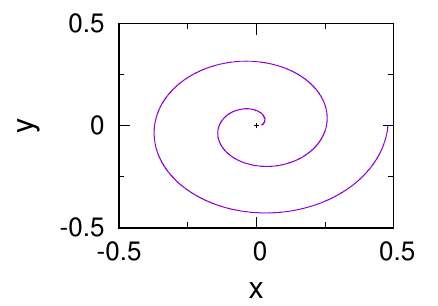}}
                \caption{Shapes for the locus (a) and its projection on to $xy$-plane (b).
                Initial values are given in figure \ref{fig:thetaphipsi_e}.
} 
        \label{fig:tt_prj_fig8j}
        \end{center}
\end{figure}

In figure \ref{fig:tt_prj_fig8j}(a), one can see the shapes for the locus for numerical results.
It can be seen that the top rises again by making a spiral shape with fewer windings.

\subsubsection{Comparison}

In this part, we will study and compare the pure slipping model, the simplified pure slipping model, i.e. equations \eqref{diffeqns_odp}, and Jellett's model.
We will consider the simplified pure slipping to get a better comparison due to the absence of air dissipation and frictional effect arising from $\dot \theta$ similar to Jellett's model.
Since the most obvious difference in the pure slipping and Jellett's models is the dependence of the rise term on $\tilde l$ in Jellett's model, we will consider a second top whose distance between the center of the peg and the center of mass is double of the previously considered top.
This can be done by changing the position of the top's disc, see figure \ref{fig:hst}.
For such a change, the moment of inertia of the symmetry axis $I_3$ does not change but moment of inertia of the transverse axis $I_1$ does.

We have already given parameters of the previously considered top, which we will name as top 1 from now on, and the one with doubled $\tilde l$ as top 2.
For the top 2, the mass, the radius of the peg and $I_z$ (and $\tilde I_z$) will be the same as the top 1, and the distance between the center of mass and the center of the peg $\tilde l_{t2}$ will be equal to $40\,mm$.
The moment of inertia about tranverse axis of top 2 $I_{t2,x}$ in the center of peg-body reference frame will be taken as $22.3\times10^{-5}\,kg\,m^2$, and $\tilde I_{t2,x}=4.64\times10^{-5}\,kg\,m^2$ in the center of mass-body reference frame.

We will use the weak top condition to get clear discrimination between the pure slipping and Jellett's models.
The weak top condition is obtained from regular precession and gives a relation between $a=I_3 (\dot \psi -\dot \phi \cos \theta)/I_1$ and $\sqrt{4 Mg \tilde l/I_1}$.
If $|a|$ is smaller than $\sqrt{4 Mg \tilde l/I_1}$, then the top is considered as weak top \cite{KleinSommerfeld}.
We should note that this discrimination works approximately when conserved angular momenta are not equal to each other.
For top 1: $\sqrt{4 Mg \tilde l_{t1}/I_{t1,x}}=31.8 \,rad\,s^{-1}$ and $\sqrt{4 Mg \tilde l_{t1}/\tilde I_{t1,x}}=46.0 \,rad\,s^{-1}$.
For top 2: $\sqrt{4 Mg \tilde l_{t2}/I_{t2,x}}=27.8 \,rad\,s^{-1}$ and $\sqrt{4 Mg \tilde l_{t2}/\tilde I_{t2,x}}=61.0 \,rad\,s^{-1}$

\begin{figure}[h!]
        \begin{center}
        \subfigure[Pure slipping]{
                \includegraphics[width=4.2cm]{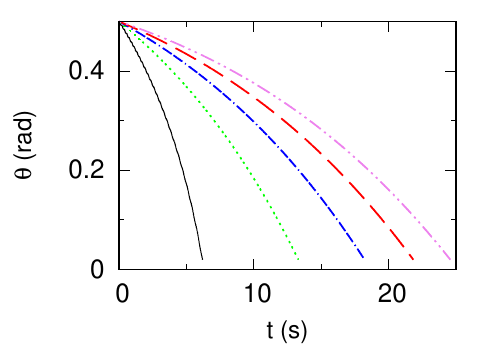}
                }
        \subfigure[Simplified pure slipping]{
                \includegraphics[width=4.2cm]{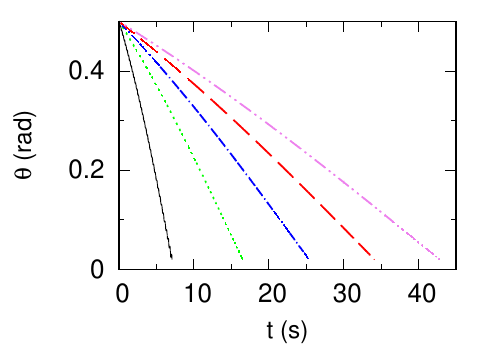}
                }
        \subfigure[Jellett's model]{
                \includegraphics[width=4.2cm]{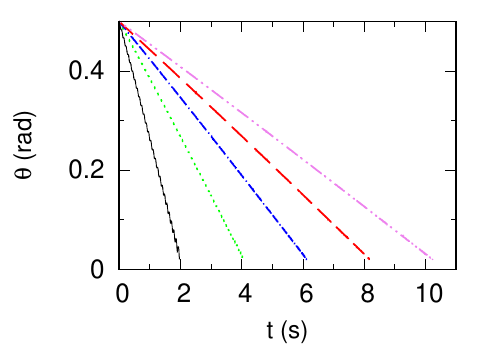}
                }
        \subfigure[Pure slipping]{
                \includegraphics[width=4.2cm]{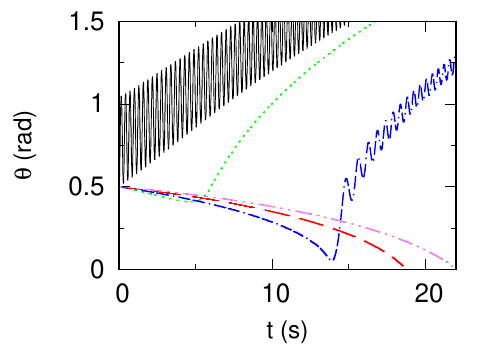}
                }
        \subfigure[Simplified pure slipping]{
                \includegraphics[width=4.2cm]{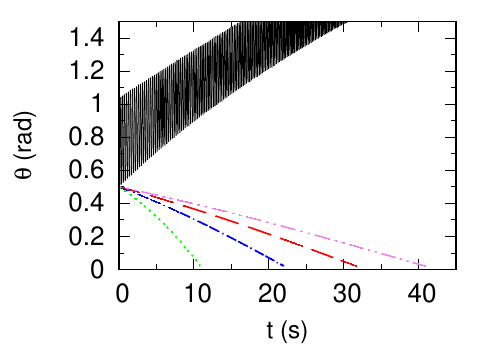}
                }
        \subfigure[Jellett's model]{
                \includegraphics[width=4.2cm]{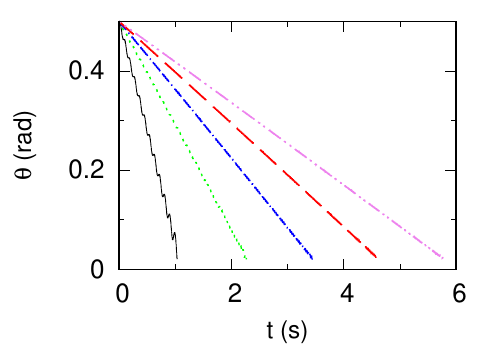}
                }
		\caption{Results of the numerical solution for $\theta$ for top 1 (upper three graphs) and 2 (lower three graphs) with different initial $\dot \psi_0$ and $\dot \phi_0$ values. 
		Continuous (black) lines show results for $\dot \psi_0=-50\,rad\,s^{-1}$,
		dotted (green) lines show results for $\dot \psi_0=-100\,rad\,s^{-1}$,
		dashed-dotted (blue) lines show results for $\dot \psi_0=-150\,rad\,s^{-1}$,
		dashed (red) lines show results for $\dot \psi_0=-200\,rad\,s^{-1}$,
		dashed-double dotted (voilet) lines show results for $\dot \psi_0=-250\,rad\,s^{-1}$.
		$\theta_0=0.5\,rad$ and $\dot \theta_0=0$ for all cases.
		$\dot \phi_0$ values are chosen very close to smaller root of $\dot \phi_{1,2}=(-I_3 \dot \psi-\sqrt{(I_3 \dot \psi)^2+4 Mg \tilde l \cos \theta})/[2 (I_3-I_1) \cos \theta]$ except the situation $\dot \psi_0=-50\,rad\,s^{-1}$ for the pure slipping and simplified pure slipping models, in which $\dot \phi_0=-35\,rad\,s^{-1}$.  
                }
        \label{fig:thetaphipsi_d2}
        \end{center}
\end{figure}

We will consider five different $\dot \psi_0$ values for both tops, and
we will numerically solve these 10 cases with mentioned models.
For all cases, $\theta_0=0.5\,rad$ and $\dot \theta_0=0$.
$\dot \psi_0$ changes from $-50 \,rad\,s^{-1}$ to $-250 \,rad\,s^{-1}$ with $50 \,rad\,s^{-1}$ decrements, and corresponding $\dot \phi_0$ values are taken as very close to the ones giving regular precession with two exceptional situations.
We should note that for the pure slipping and simplified pure slipping models, $\dot \phi_0$ are taken as the same since $I_x$ are the same for these models, and $\dot \phi_0$ is slightly different for Jellett's model since $\tilde I_x$ is different.
In exceptional situations, $\dot \psi=-50 \,rad\,s^{-1}$ for the pure slipping and simplified pure slipping models for top 2, the top is weak and regular precession is not possible, and for these two situations, $\dot \phi_0$ is taken as $-35.0 \,rad\,s^{-1}$.
One can see results of numerical solutions for $\theta$ for mentioned 30 situations in figure \ref{fig:thetaphipsi_d2}.

The upper three graphs of the figure \ref{fig:thetaphipsi_d2} show results for top 1 with the order the pure slipping, the simplified pure slipping and Jellett's models, 
and the lower three graphs show corresponding ones for top 2.
It can be seen that as the magnitude of $\dot \psi_0$ increases the rise time increases which arises from the need for greater $\dot \phi$ values for the rise.
The same thing also explains the difference in rise times between the pure slipping and simplified slipping models since air dissipation helps the decrease of $\dot \psi$ in the pure slipping.
The dependence of the rise term on $\tilde l$ in Jellett's model causes a faster rise of tops in all cases.

In previously mentioned exceptional two situations, top 2 does not rise and fall.
On the other hand, top 2 rises in the corresponding situation when the solution is obtained by using Jellett's method.
For mentioned two situations, $|a|=26.2\,rad\,s^{-1}$ which is smaller than $27.8 \,rad\,s^{-1}$ in the pure slipping and simplified pure slipping models; but in Jellett's model, $|a|=93.3 \,rad\,s^{-1}$ which is greater than $61.0 \,rad\,s^{-1}$.
Then, this case gives weak top for the pure slipping and simplified pure slipping models, which results in fall.
On the other hand, this situation is not weak in the center of mass-body reference frame, and top 2 rises in the corresponding situation in Jellett's model.

Another remarkable thing is seen when top 2 is not weak and not strong enough for the whole rise in the pure slipping model.
It can be seen from figure \ref{fig:thetaphipsi_d2}(d) that top 2 does not completely rise in the pure slipping model when $\dot \psi_0=-100 \,rad\,s^{-1}$ and $\dot \psi_0=-150 \,rad\,s^{-1}$.
The top rises at the beginning, and during this process, its spin angular velocity decreases mainly due to air dissipation, and it starts to fall around $t=4.9s$ and $t=13.9s$, respectively,
and numerical results show that in these cases, top 2 becomes weak at $t=4.7s$ and $t=13.7s$, respectively.
On the other hand, it rises entirely in Jellett's model and simplified pure slipping.
The reason for the rise in simplified pure slipping is the absence of air dissipation.

When top 2 is strong enough for the whole rise in the pure slipping model, it is possible to compare rise times of different models. 
The percentage differences in the rise time are 15\% are 11\% for the last two cases in the pure slipping model, they are 7\% and 4\% in the simplified pure slipping model, and they are 43\% in Jellett's model.

\section{Conclusion}
\label{five}

We have studied dissipative motion and the rise of the heavy symmetrical top with a hemispherical peg.
Differently from previous works, we studied motion by considering the radial center of the hemispherical peg as the fixed point of the top.
From the derivation, we obtained the rise term as a function of the radius of the peg $R_p$, which is consistent with the experiments \cite{Fokker1941}.
On the other hand, in Jellett's model \cite{Jellett}, and in other similar ones \cite{Parkyn, Yogi, Moffatt}, the center of mass is considered as the fixed point and the rise term is a function of $\tilde l$ together with $R_p$. 

The difference between the rise terms provides the possibility of determination of the better model with simple experiments.
By using a top with a changeable disc position, it is possible to change the position of the center of mass which can provide necessary experimental data.
 
We should note that we have determined dissipative constants according to the pure slipping model by considering an experimentally studied case.
There should be some differences in these constants if one had used any other model.
However, the differences between Jellet's model and the pure slipping model are bigger, and any change in the dissipative constants does not substantially affect the following comments.

One can use strong tops and compare changes in rise times and decide the better model.
By looking at the difference between the simplified pure slipping and pure slipping models, one can say that if one includes air dissipation and friction due to $\dot \theta$ at the touchpoint in Jellett's model, the rise time should decrease.
However, this decrease should be smaller than the difference between the simplified pure slipping and pure slipping models, since rise time in Jellett's model is smaller.
By considering the results of numerical solutions, one can say that after doubling the distance between the center of mass and the center of the peg, the rise time should decrease.
For strong tops, if the rise time changes around 15\% then the pure slipping model is better; 
on the other hand, if the rise time changes around 40\% then Jellett's model is better.
We should note that the friction coefficient and the radius of the peg should be convenient to get a longer rise time which may provide better data for discrimination.

Obviously, one can use the weak top condition to decide the better model.
If one makes an experiment with a case that is weak in center of peg-reference frame and strong in center of mass-reference frame, then one can easily decide the better model:
If the top rises, then Jellett's model is better; and if it falls, then the pure slipping model is better.

\section{Appendix}

\begin{figure}[h!]
        \begin{center}
                \subfigure[$\theta$]{\includegraphics[width=4.2cm]{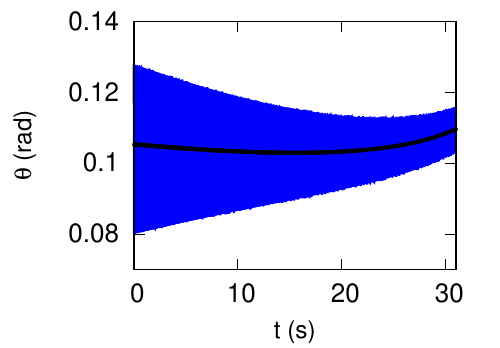}}
                \subfigure[$\dot \theta$]{\includegraphics[width=4.2cm]{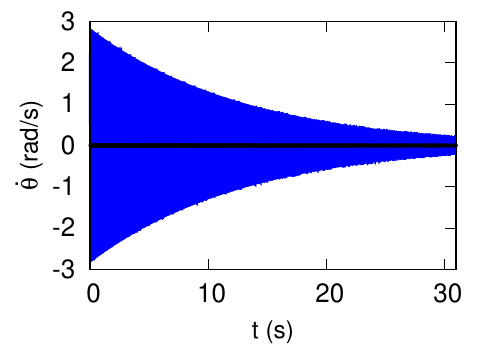}}

                \subfigure[$\dot \phi$]{\includegraphics[width=4.2cm]{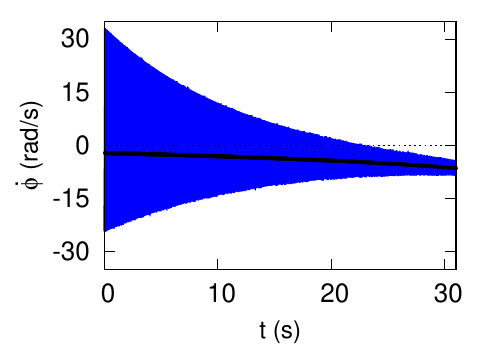}}
                \subfigure[$\dot \psi$]{\includegraphics[width=4.2cm]{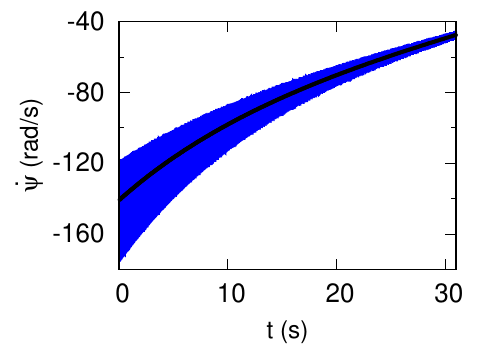}}
        \caption{Results of the numerical solution for the dissipative motion for $\theta$ (a), $\dot \theta$ (b), $\dot \phi$ (c) and $\dot \psi$ (d). Continuous (blue) lines show results of numerical solution and dots (black) show nutation average.
                Initial values are the same as figure \ref{fig:prj_fig5}. }
        \label{fig:thetaphipsi_c}
        \end{center}
\end{figure}

Figure \ref{fig:thetaphipsi_c} shows results of numerical solution for dissipative motion for $\theta$ and angular velocities.

\end{document}